\documentclass{webofc}
\usepackage[varg]{txfonts}   
\def\be{\begin{equation}} 
\def\ee{\end{equation}} 
\def\Tr{{\rm Tr}} 
\begin{document}
\title{Magnetic dipole $\gamma$-ray strength functions of heavy nuclei in the configuration-interaction shell model}
%
%

\author{\firstname{Y.} \lastname{ Alhassid}\inst{1}\fnsep\thanks{\email{yoram.alhassid@yale.edu}} \and
        \firstname{P.} \lastname{ Fanto}\inst{1}\fnsep\thanks{\email{paul.fanto@yale.edu}} \and
        \firstname{A.}\lastname{ Mercenne}\inst{1,2}\fnsep\thanks{\email{amercenne@lsu.edu} 
        }
}
\institute{Center for Theoretical Physics, Sloane Physics Laboratory,
 Yale University, New Haven, Connecticut  06520, USA 
\and  Department of Physics and Astronomy, Louisiana State University, Baton Rouge, Louisiana 70803, USA   
          }

 \abstract{
 A low-energy enhancement (LEE) has been observed in the deexcitation $\gamma$-ray strength function ($\gamma$SF) of compound nuclei. The LEE has been a subject of intense experimental and theoretical interest since its discovery, and, if the LEE persists in heavy neutron-rich nuclei, it would have significant effects on calculations of r-process nucleosynthesis.  Standard configuration-interaction (CI) shell-model calculations in medium-mass nuclei have attributed the LEE to the magnetic dipole $\gamma$SF but such calculations are computationally intractable in heavy nuclei.  We review a combination of beyond-mean-field many-body methods within the framework of the CI shell model that enables the calculation of $\gamma$SF in heavy nuclei, and discuss the recent theoretical identification of a LEE in the magnetic dipole $\gamma$SF of lanthanide isotopes.}
\maketitle
\section{Introduction}\label{intro}

$\gamma$-ray strength functions ($\gamma$SF)~\cite{barth1973}, along with level densities, are necessary input to the Hauser-Feshbach theory~\cite{hauser1952} of compound nucleus reactions.  In recent years, a low-energy enhancement (LEE) was observed in the $\gamma$SFs of mid-mass nuclei~\cite{voinov2004,larsen2013,larsen2018} and in several rare-earth nuclei~\cite{simon2016,naqvi2019,guttormsen2022}.  If the LEE persists in heavy neutron-rich nuclei, it would have significant effects on the calculations of $r$-process nucleosynthesis by enhancing radiative neutron capture rates near the neutron drip line~\cite{larsen2010}.

Conventional configuration-interaction (CI) shell-model calculations have attributed the LEE to the magnetic dipole ($M1$) $\gamma$SF~\cite{schwengner2013, brown2014, karampagia2017, schwengner2017, midtbo2018}. However, CI shell model calculations are prohibited in heavy nuclei because of the combinatorial increase in the dimensionality of the many-particle model space with  increasing number of valence orbitals and/or number of valence nucleons. 

The shell-model Monte Carlo (SMMC) method~\cite{johnson1992, alhassid1994, koonin1997, alhassid2008} enables microscopic calculations in model spaces  that are many orders of magnitude larger than those that can be treated by conventional diagonalization methods; see Ref.~\cite{alhassid_rev} for a recent review.  It has been established as the state-of-the-art method for the microscopic calculation of level densities. 

The microscopic calculation of $\gamma$SFs in the framework of SMMC has been a major challenge. A strength function is usually calculated from the Fourier transform of the real-time response function. However, in SMMC we can only calculate
imaginary-time response functions whose inverse Laplace transform is the strength function. This inverse Laplace
transform is numerically an ill-posed problem and requires the analytic continuation to real time. This can be done numerically using
the maximum entropy method~\cite{gubernatis1991, jarrell1996,gubernatis_book}, but its success depends crucially on a good choice of a prior strength function.
We found that the static-path approximation plus random-phase approximation (SPA+RPA)~\cite{puddu1991, attias1997, rossignoli1998b} provides a good prior strength function~\cite{fanto_arXiv}. The SPA itself also provides a good prior strength at not too low temperatures~\cite{mercenne2023}. The SPA+RPA includes large-amplitude static fluctuations~\cite{lauritzen1988} of the auxiliary fields plus small time-dependent fluctuations around each static configuration.  It was recently applied to calculate state densities in heavy nuclei in close agreement with SMMC state densities~\cite{fanto2021}.  

\section{The shell model Monte Carlo (SMMC) method}

SMMC is based on the Hubbard-Stratonovich transformation~\cite{stratonovich1957, hubbard1959}, in which the Gibbs ensemble $e^{-\beta \hat H}$ is written as a superposition of ensembles $U_\sigma$ of non-interacting nucleons that move in time-dependent fields $\sigma(\tau)$
\begin{equation}
\label{HS}
e^{-\beta H} = \int D[\sigma]  G_\sigma U_\sigma \;,
\end{equation}
where $G_\sigma$ is a Gaussian weight and $D[\sigma]$ is an integration measure.

The calculation of the integrand in Eq.~(\ref{HS}) reduces to matrix algebra in the single-particle space of typical dimension $\sim 50 - 100$ for heavy nuclei. However, the main challenge is carrying out the integration over the large number of auxiliary fields. This is done through the use of Monte Carlo methods to select the important configurations of  the auxiliary fields.

SMMC is a powerful method to calculate thermal observables in the presence of correlations but it cannot be used to calculate directly $\gamma$SFs.  In SMMC, we can calculate the imaginary-time response function of the respective electromagnetic operator $\hat O_\lambda$, and it is necessary to carry out the analytic continuation to real time.

\section{Finite-temperature $\gamma$SF}

The finite-temperature strength function of an hermitean tensor operator $O_\lambda$ of rank $\lambda$ (e.g., $E1$, $M1$,\ldots) is defined by
\begin{eqnarray}\label{strength_cism}
S_{\!\mathcal{O}_\lambda}(\omega) = \sum_{\substack{\alpha_i J_i \\ \alpha_f J_f}} \frac{e^{-\beta E_{\alpha_i J_i}}}{Z} |(\alpha_f J_f || \hat{\mathcal{O}}_\lambda|| \alpha_i J_i  )|^2 \delta(\omega - E_{\alpha_f J_f}+ E_{\alpha_i J_i })\,,
\end{eqnarray}
where $(\alpha J)$ denotes eigenstates with energy $E_{\alpha J}$ and spin  $J$, and $Z = \sum_{\alpha J} (2J+1) e^{-\beta E_{\alpha J}}$ is the canonical partition function.  
The strength function can be written as a Fourier transform of the real-time response function of $\hat O_\lambda$. However, in SMMC it is only possible to calculate the imaginary-time response function
\begin{equation}
R_{\mathcal{O}_\lambda} (\tau) =  \left\langle  \sum_{\lambda \mu} (-)^\mu \hat O_{\lambda \mu} (\tau) \hat O_{\lambda -\mu}(0) \right\rangle \;,
\end{equation}
where the expectation value is over the canonical ensemble at temperature $T$ and $\hat O_{\lambda \mu}(\tau) = e^{\tau \hat H} \hat O_{\lambda \mu} e^{-\tau \hat H}$. The strength function is then the Laplace transform of the imaginary-time response function. The $\gamma$SF satisfies the symmetry relation $ S_{\!\mathcal{O}_\lambda}(-\omega)= e^{-\beta \omega} S_{\!\mathcal{O}_\lambda}(\omega)$ and the Laplace transform can be rewritten in the form
\begin{equation}\label{kernel}
R_{\mathcal{O}_\lambda}(\tau) = \int_0^\infty d\omega\, K(\tau,\omega) S_{\!\mathcal{O}_\lambda}(\omega)\,,
\end{equation}
where $K(\tau,\omega)= e^{-\tau\omega} + e^{-(\beta-\tau) \omega}$ is a symmetrized kernel.

\section{Analytic continuation: maximum-entropy method}\label{MEM}

The inversion of (\ref{kernel}) to calculate the strength function from the imaginary-time response function requires analytic continuation to real time and is numerically an ill-defined problem that does not have a unique solution. We use the maximum entropy method (MEM)~ \cite{jarrell1996} to carry out this analytic continuation numerically.  In this method we choose the strength function that maximizes the objective function 
 \begin{equation}  \label{maxent}
   \mathcal{Q}(S_{\!\mathcal{O}_\lambda}; \alpha) = \alpha \mathcal{S} - \frac{1}{2}\chi^2\,
  \end{equation}
  at a given value of $\alpha$.  Here $\chi^2$ is given by
 \begin{equation}
\chi^2 = (\overline{R}_{\mathcal{O}_\lambda} - R_{\mathcal{O}_\lambda})^T \mathcal{C}^{-1} (\overline{R}_{\mathcal{O}_\lambda} - R_{\mathcal{O}_\lambda})\,,
  \end{equation}
  where ${ \bar{R} }$ and ${ \mathcal{C} }$ are the SMMC response function and its covariance matrix, respectively. $\mathcal{S}$ is the entropy function
\begin{equation}\label{entropy}
\mathcal{S} =  -\int d\omega\, \bigg( S_{\!\mathcal{O}_\lambda}(\omega) - S^{\rm prior}_{\!\mathcal{O}_\lambda}(\omega)
- S_{\!\mathcal{O}_\lambda}(\omega)\ln \left[S_{\!\mathcal{O}_\lambda}(\omega)/S^{\rm prior}_{\!\mathcal{O}_\lambda}(\omega)\right]\bigg)\,,
\end{equation}
where $S^{\rm prior}_{\!\mathcal{O}_\lambda}(\omega)$ a prior strength function. The coefficient $\alpha$ controls the relative weight between the entropy and the $\chi^2$ terms.  We use Bryan's method~\cite{bryan1990}, in which the final strength function is obtained by averaging over $\alpha$ with a certain probability distribution.

\section{Prior strength function: SPA + RPA}

The choice of a good prior strength is key to the success of the MEM. We evaluate the prior strength function in the SPA+RPA~\cite{puddu1991, attias1997, rossignoli1998b} 
\begin{equation}\label{spa_rpa_gsf}
S^{\rm prior}_{\!\mathcal{O}_\lambda}(\omega) \approx \frac{\int d\sigma M(\sigma) Z_\eta(\sigma) C_\eta(\sigma) S_{\!\mathcal{O}_\lambda, \eta}(\sigma;\omega)}{\int d\sigma M(\sigma) Z_\eta(\sigma) C_\eta(\sigma)}\,,
\end{equation}
where $\sigma$ are static auxiliary fields,  and $M(\sigma)$ is a measure~\cite{fanto2021}. The quantity $Z_\eta(\sigma) = \Tr\left[ \hat P_\eta e^{-\beta \left(\hat h_\sigma - \sum_{\lambda=p,n} \mu_\lambda \hat N_\lambda\right)}\right]$ is the number-parity projected one-body partition function, where $\hat h_\sigma$ is a one-body Hamiltonian for a given static configuration of the fields, and $P_\eta=(1+\eta e^{i\pi \hat N})/2$ is the number-parity projection with $\eta = +1(-1)$ for even (odd) number parity.  $C_\eta(\sigma)$ is the RPA correction factor that accounts for the Gaussian integration over small amplitude time-dependent auxiliary-field fluctuations~\cite{puddu1991, attias1997, rossignoli1998a, nesterov2013, fanto2021}. $S_{\!\mathcal{O}_\lambda, \eta}(\sigma;\omega)$ is a $\sigma$-dependent QRPA strength function at temperature $T$; see Eqs.~(5) and (6) in Ref.~\cite{fanto_arXiv}. The integration in Eq.~(\ref{spa_rpa_gsf}) is carried out using the Metropolis-Hastings algorithm with a weight function of $\mathcal{W}_\eta(\sigma) = M(\sigma) Z_\eta(\sigma)$.

The above SPA+RPA strength function was used as prior for a pairing plus quadrupole interaction~\cite{fanto_arXiv}.  The complete SMMC interaction includes additional components, and the SPA+RPA, which requires the diagonalization of the QRPA matrix at many static configurations of the auxiliary field, becomes time consuming.  Instead we apply the SPA alone~\cite{mercenne2023}, for which $C_\eta(\sigma)=1$ and the strength function in the integrand is the $\sigma$-dependent SPA strength function.

 \section{$M1$ $\gamma$SFs in lanthanide isotopes}
 
 We applied the methods discussed above to isotopic chains of lanthanides.  The single-particle valence model space (using Woods-Saxon central potential plus spin-orbit interaction) is the 50-82 shell plus the $1f_{7/2}$ orbital for protons, and the 82-126 shell plus the $0h_{11/2}$ and $1g_{9/2}$ orbitals for neutrons. For the samarium isotopes we used a pairing plus quadrupole interaction and for the neodymium isotopes we used monopole pairing and multipole-multipole interaction that includes quadrupole, octupole and hexadecupole  components~\cite{ozen2013}.

For the magnetic dipole transition operator $M1$ we used
\begin{equation}\label{M1}
\hat{\mathcal{O}}_{M1} = \sqrt{\frac{3}{4\pi}} \frac{\mu_N}{\hbar c} \left( g_l \mathbf{l} + g_s \mathbf{s}\right)\,,
\end{equation}
where $\mathbf{l}$ and $\mathbf{s}$ are the orbital and spin angular momentum operators, respectively.  We used the free-nucleon $g$ factors $g_{l,p} = 1$, $g_{l,n} = 0$, $g_{s,p} = 5.5857$, and $g_{s,n} = -3.8263$.

\subsection{Samarium isotopes}

\begin{figure*}[h!]
\centering
\includegraphics[width=12.5 cm,clip]{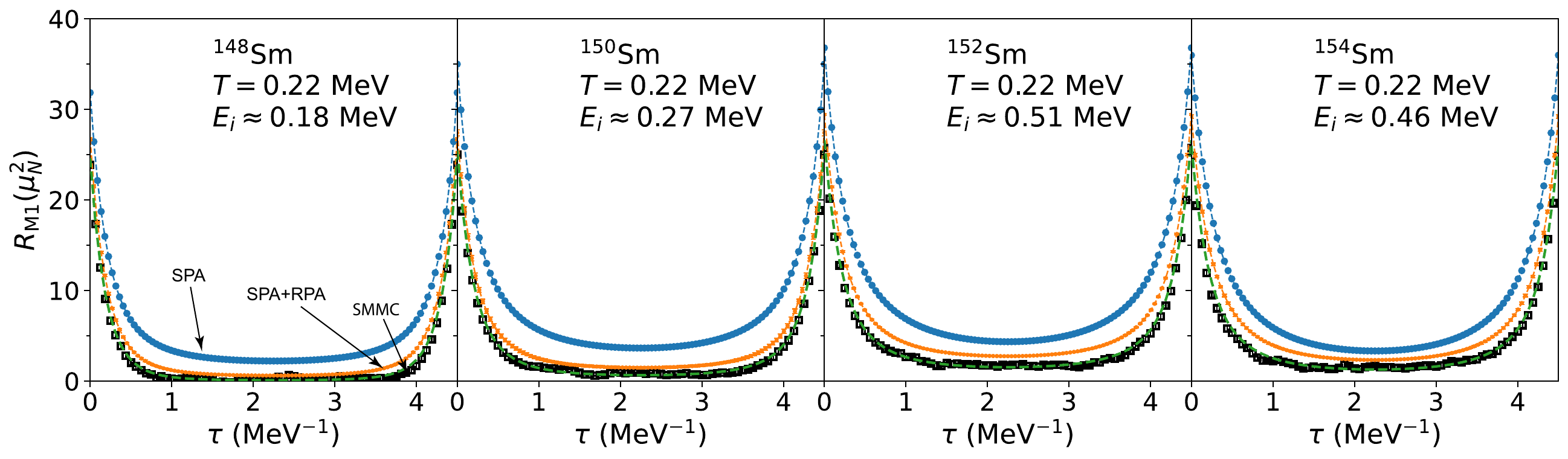}
\caption{The $M1$ imaginary-time response functions in a chain of even-mass samarium isotopes. The SPA +RPA results (orange dots) are compared with the SMMC results (black dots). Also shown are the SPA (blue circles) and MEM (green dashed lines) response functions. Adapted from Ref.~\cite{fanto_arXiv}.}
\label{response}     
\end{figure*}

In Fig.~\ref{response} we show the imaginary-time $M1$ response function for chain of samarium isotopes at a low temperature of $T=0.22$ MeV. The SPA+RPA response function (orange dots) is close to the exact SMMC response function (black squares), indicating that the SPA+RPA is a good choice for a prior $M1$ $\gamma$SF.  For comparison we also show the SPA response function (blue circles).

\begin{figure*}[h!]
\centering
\includegraphics[width=12.5 cm,clip]{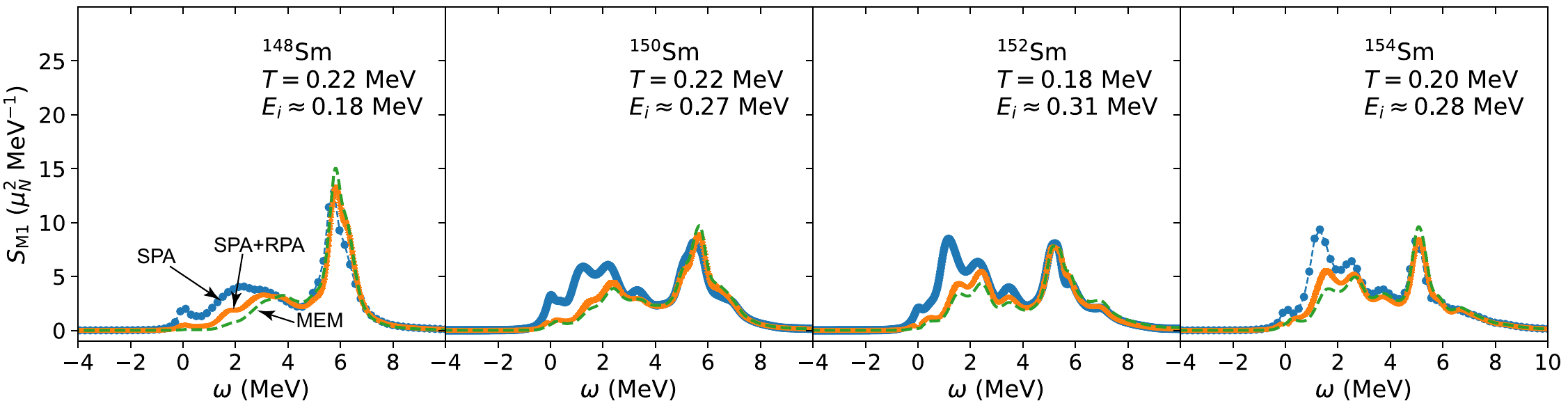}
\caption{$M1$ $\gamma$SFs near the ground state in a chain of even-mass samarium isotopes.  The MEM $M1$ $\gamma$SFs (dashed green lines)  are compared with the SPA (blue circles) and the SPA+RPA (orange dots) strengths. Adapted from Ref.~\cite{fanto_arXiv}.}
\label{strength_gs_Sm}     
\end{figure*}

The corresponding $M1$ $\gamma$SFs are shown in Fig.~\ref{strength_gs_Sm} with positive values of $\omega$ describing absorption of $\gamma$-rays. We apply the MEM discussed in Sec.~\ref{MEM} with the SPA+RPA strength as the prior strength to find the dashed green lines.  These provide us with the $M1$ $\gamma$SFs near the ground state. We observe a peak around 6 MeV, which we interpret as the spin-flip mode. As we add neutrons and the nucleus becomes more deformed, some of the strength in this peak is transferred to a structure around 1-3 MeV, which is consistent with the scissors mode known in heavy deformed nuclei~\cite{bohle1984,heyde2010}. 

\begin{figure*}[h!]
\centering
\includegraphics[width=12.5 cm,clip]{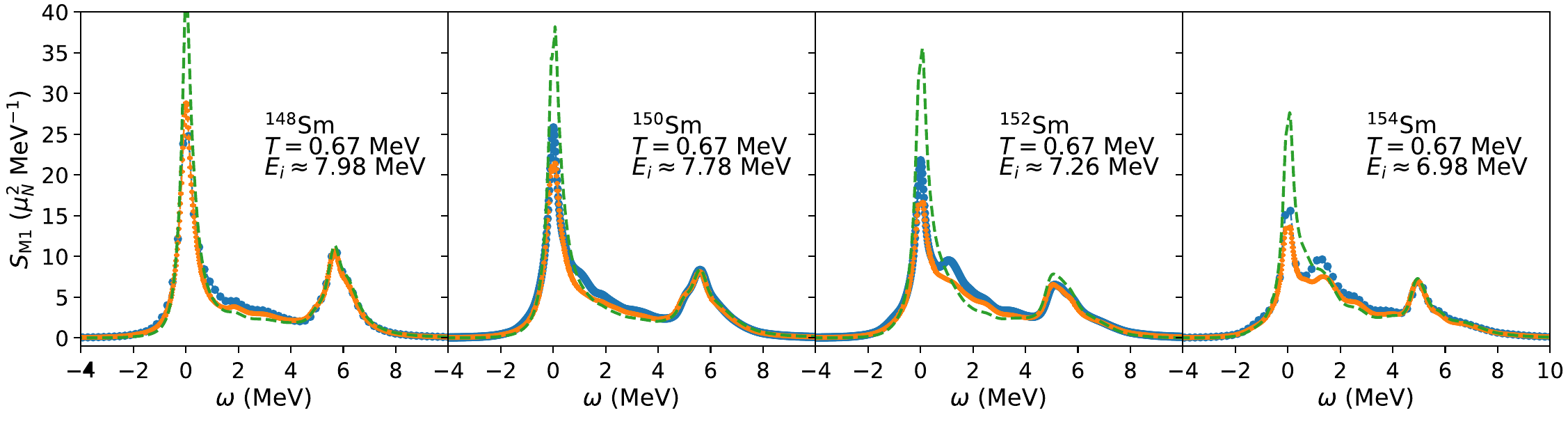}
\caption{Same as in Fig.~\ref{strength_gs_Sm} but at temperatures that correspond to the neutron resonance energies.
Adapted from Ref.~\cite{fanto_arXiv}.}
\label{strength_res_Sm}     
\end{figure*}

In Fig.~\ref{strength_res_Sm} we show the corresponding $M1$ $\gamma$SFs for the samarium isotopes but at temperatures that correspond to the neutron resonance energies.  We observe a pronounced peak at $\omega \approx 0$ that, as shown below, leads to a low-energy enhancement in the $\gamma$-ray decay.  As neutron number increases, the strength of this peak reduces and a small peak develops at $\omega \approx 2$ MeV, which we interpret as a scissors mode built on top of excited states~\cite{krticka2011}.  We still observe the spin-flip mode at $\omega \approx 6$ MeV.

\subsection{Neodymium isotopes}

\begin{figure*}[h!]
\centering
\includegraphics[width=13. cm,clip]{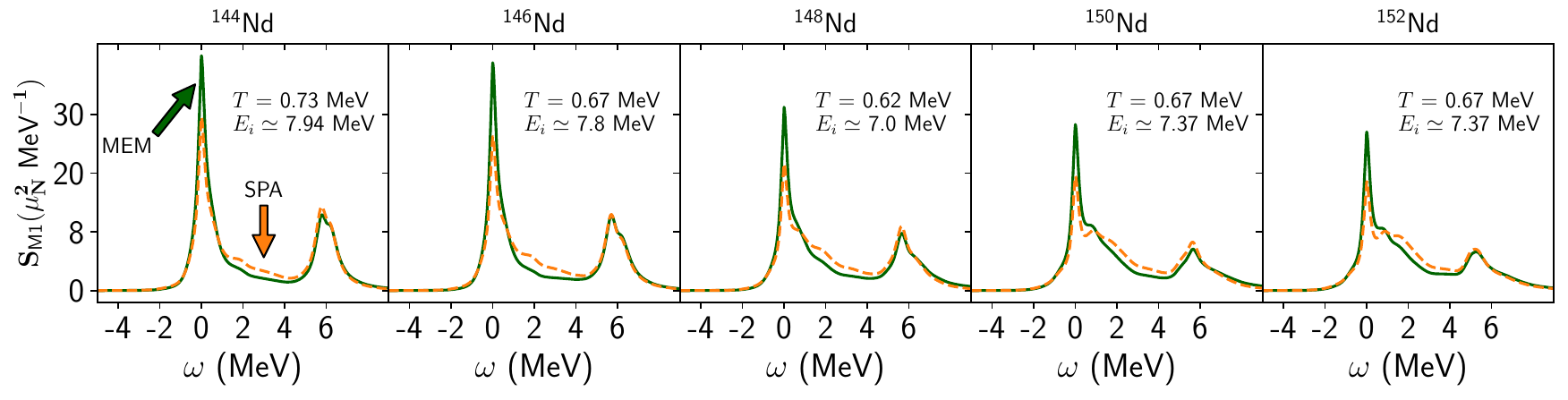}
\caption{$M1$ $\gamma$SFs at the corresponding neutron resonance energies for a chain of even-mass neodymium isotopes. The MEM $\gamma$SFs (green lines) are compared with the SPA $\gamma$SFs (dashed orange lines). Adapted from Ref.~\cite{mercenne2023}.}
\label{strength_res_Nd}     
\end{figure*}

We calculated the $M1$ $\gamma$SFs for a chain of even-mass neodymium isotopes, for which recent experimental data are available. We used the SPA strength as the prior strength for the MEM.  In Fig.~\ref{strength_res_Nd} we show the corresponding $M1$ $\gamma$SFs for the neodymium isotopes at temperatures that correspond to the neutron resonance energies. The MEM results (green lines) are compared with the SPA strengths (dashed orange lines). 

We observe similar features as for the samarium isotopes, and in particular a peak at $\omega \approx 0$. We also identify a spin-flip mode close to 6 MeV, and with the addition of neutrons a scissors mode around 1-2 MeV, both of which are built on top of excited states.

Experimental studies often use the deexcitation strength function $f_{M1}$ which can be related to $S_{\!M1}$ by~\cite{private}
\begin{equation}
      {f}_{M1}({E}_{\gamma}, {E}_{i}) \approx \frac{1}{3}a \frac{\tilde\rho(E_i)}{\tilde\rho(E_i-E_\gamma)} {S}_{M1}(\omega = - {E}_{\gamma}) \;,
      \label{eq_gsf}
  \end{equation}
   where $a=\frac{16 \pi}{9(\hbar c)^3}$ and ${E}_{\gamma}$ is the emitted $\gamma$-ray energy. $\tilde\rho(E_i)$ and $\tilde\rho(E_i-E_\gamma)$ are the initial and final level densities, respectively. 
   
\begin{figure*}[h!]
\centering
\includegraphics[width=13. cm,clip]{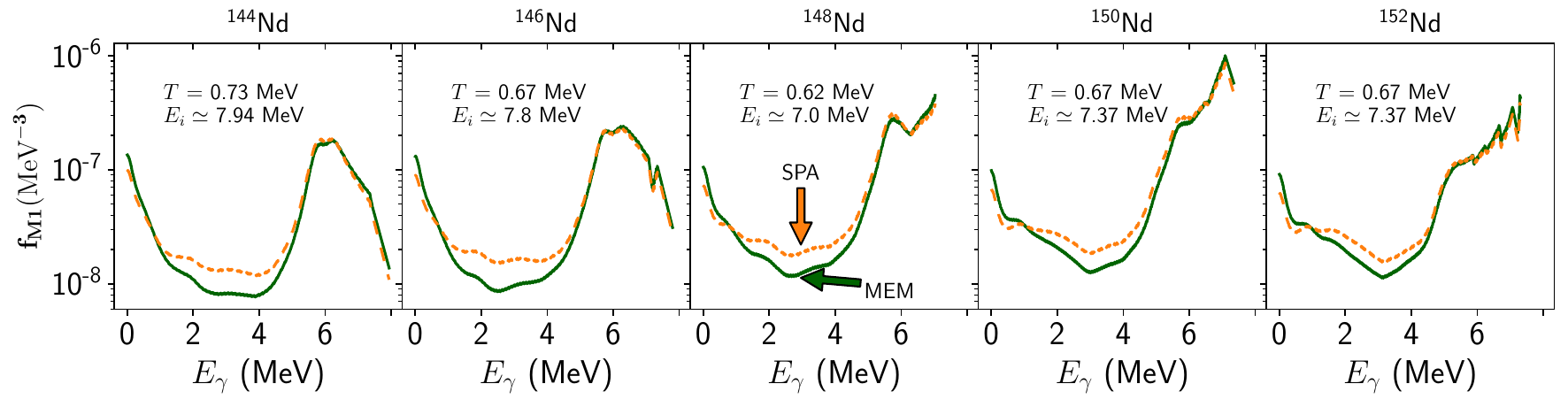}
\caption{$f_{M1}$ vs.~$E_\gamma$ in a chain of even-mass neodymium isotopes. The MEM results (green lines) are compared with the SPA results (dashed orange lines). Adapted from Ref.~\cite{mercenne2023}.}
\label{f_res_Nd}     
\end{figure*}

 In Fig.~\ref{f_res_Nd} we show $f_{M1}$ for the even-mass neodymium isotopes as a function of the emitted $\gamma$-ray energy $E_\gamma$. We used Eq.~(\ref{eq_gsf}), where the level densities of the neodymium isotopes were computed within SMMC using spin projection methods~\cite{guttormsen2021}. We clearly observe a LEE at low $\gamma$-ray energies that originates from the $\omega \approx 0$ peak in $S_{\! M1}(\omega)$. 
 
 We find that the LEE follows an exponential decay $\sim e^{-E_\gamma/T_B}$ as was observed in Refs.~\cite{schwengner2013, karampagia2017}.  The fitted values of $T_B$ are approximately constant over a range of initial energies. 

\section{Conclusion}

We reviewed a  method to calculate $\gamma$-ray strength functions for heavy nuclei in the the framework of the configuration-interaction shell model by combining two many-body methods: SMMC and the SPA+RPA or SPA~\cite{fanto_arXiv,mercenne2023}.  Using this method, we made the first theoretical identification of a low-energy enhancement (LEE) in heavy nuclei. In particular, we observed a LEE in the $M1$ $\gamma$SFs in isotopic chains of samarium~\cite{fanto_arXiv} and neodymium~\cite{mercenne2023} nuclei. We also observed the emergence of a structure consistent with a scissors
 mode built on top of excited states as neutron number increases within each chain.

The LEE was observed experimentally in lanthanide isotopes~\cite{simon2016,naqvi2019,guttormsen2022} in experiments based on the Oslo method, but these experiments cannot separate the $E1$ and $M1$ components of the $\gamma$SF.   A detailed comparison with such experiments requires the calculation of the $E1$ $\gamma$SF. 

\section{Acknowledgments}
This work was supported in part by the U.S. DOE grant No.~DE-SC0019521, and by the U.S. DOE NNSA Stewardship Science Graduate Fellowship under cooperative agreement No.~NA-0003960. The calculations used resources of the National Energy Research Scientific Computing Center (NERSC), a U.S. Department of Energy Office of Science User Facility operated under Contract No.~DE-AC02-05CH11231.  We thank the Yale Center for Research Computing for guidance and use of the research computing infrastructure.

\end{document}